\documentclass[preprintnumbers,amsmath,amssymb,showpacs,twocolumn]{revtex4}

\usepackage{graphicx}
\usepackage{dcolumn}
\usepackage{bm}

\begin{document}

\title{ A  scheme for quantum  communication using EPR pairs and local measurement}

\author{ Feng-Li Yan $^{1,2}$, Ting Gao $^{1,3,4}$}

\affiliation {$^1$ CCAST (World Laboratory), P.O. Box 8730, Beijing 100080, China\\
$^2$  College of Physics, Hebei Normal University, Shijiazhuang 050016, China\\
$^3$ College of Mathematics and Information Science, Hebei Normal University, Shijiazhuang 050016, China\\
$^4$ Department of Mathematics, Capital Normal University, Beijing 100037, China}

\begin{abstract}
 We present a  scheme for quantum  communication, where a set of EPR
pairs, initially shared by the sender Alice and the receiver Bob,
functions as a quantum  channel. After insuring the safety of the
quantum channel, Alice applies local measurement on her particles
of the EPR pairs and informs Bob the encoding classical
information publicly. According to  Alice's  classical information
and his measurement outcomes on the EPR pairs Bob can infer the
secret messages directly. In this scheme, to transmit one bit
secret message, the sender Alice only needs to send one bit
classical information to the receiver Bob. We also show that this
scheme is completely secure if perfect quantum channel is used.
\end{abstract}


\maketitle

\section{Introduction}

Cryptography is the art of enabling two parties to communicate in
private. It plays an increasing important role in the whole world.
Before transmitting their secret messages the two distant parties
must distribute secret key first. As it is unsafe  to distribute
secret key through a classical channel, people have paid a lot of
attention to quantum key distribution, the most advanced
application of the principles of quantum mechanics such as the
uncertainty principle and quantum correlations. Quantum key
distribution or quantum cryptography provides a secure way for two
remote parties, say Alice and Bob, to create a randomly binary
string that can be used as a secret key with which they can
communicate securely using Vernam one-time pad crypto-system. In
1984, Bennett and Brassard presented the first quantum
cryptography,  BB84 protocol \cite {s1}. Ekert proposed another
quantum key distribution scheme depending on the correlation of
Einstein-Podolsky-Rosen (EPR) pair, the maximally entangled state
of two particles in 1991 \cite {s2}. Afterward Bennett  also put
forward a quantum cryptography scheme known as B92 protocol \cite
{s3}. Up to now, there have already been a lot of works in quantum
key distribution on both theoretical and experimental aspects [4 -
18].

Recently,  novel quantum secure direct communication protocols
were proposed by Shimizu and  Imoto [19, 20] and Beige  et al.
\cite {s21}. In these protocols,  the two parties communicate
important messages directly without first establishing a shared
secret key to encrypt them and the messages are deterministically
sent through the quantum channel, but can only be decoded after a
final transmission of classical information. A direct
communication scheme, the "ping pong protocol" which is insecure
if it is operated in a noisy quantum channel, as indicated by
W$\acute{\rm o}$jcik \cite {s22}, was put forward by Bostr\"{o}m
and Felbinger \cite {s23}. More recently Deng et al. gave a
two-step quantum direct communication scheme using EPR pair block
\cite {s24} and a secure direct communication protocol with a
quantum one-time-pad \cite {s25}. However, in all these secure
direct communication schemes it is necessary to send the qubits
with secret messages in the public channel. Therefore a potential
eavesdropper, Eve, can attack the qubits in transmission. In order
to prevent the qubits transmitted in the public channel, we
suggested a scheme for secure direct communication between Alice
and Bob, using EPR pairs  and teleportation \cite {s26},  and two
quantum secure direct communication protocols, one by EPR pairs
and entanglement swapping \cite {s27} and the other with GHZ
states and entanglement swapping \cite {s28}.   Gao et al.
provided two schemes for controlled and secure direct
communication using three-particle entangled state and
teleportation [29, 30]. Since there is not a transmission of the
qubits carrying the secret messages between two communication
parties in the public channel, they are secure for direct secret
communication if perfect quantum channel is used. In the protocol
of Ref. \cite {s26}, for transmitting one bit secret message,
Alice have to send two bits classical information to Bob, because
the Bell measurement would produce four random outcomes, so  it
would waste the classical information resource.

In this paper, we would  like to improve the quantum communication
scheme stated in Ref. \cite {s26} and give a
 simpler but more economical  one. In the new quantum communication scheme Alice only requires to send  one
bit classical information to Bob for her to transmit one bit
secret message.

\section{Scheme for quantum  communication}

 We suppose that the two communication parties Alice and Bob share   a set of EPR pairs, the
maximally entangled pair  in the Bell state
\begin{equation}
|\Phi^+\rangle_{AB}=\frac {1}{\sqrt
2}(|00\rangle_{AB}+|11\rangle_{AB}).
\end{equation}
As a matter of fact, there are many different ways for Alice and
Bob to obtain these EPR pairs. For instance, Alice makes the pairs
first then sends half of each to Bob. Or a sever could prepare the
pairs and then send half of each to Alice and Bob. In order to
make sure of the purity of EPR pairs, Alice and Bob must do some
tests. They can use the schemes testing the security of EPR pairs
(quantum channel) in Refs.[2, 4, 13, 24, 26]. Passing the test
insures that they continue to hold sufficiently pure, entangled
quantum states. However, if tampering has occurred, they discard
these EPR pairs and  construct new EPR pairs again.

In virtue of these pure EPR pairs, Alice and Bob can start their
quantum communication. Suppose that Alice wishes to communicate to
Bob. First Alice makes the local measurement on her particle A in
the basis $\{|0\rangle_A, |1\rangle_A\}$. Alice and Bob agree on
that if Alice's measurement outcome is the same with the
 secret message to be transmitted, then Alice sends classical information 0 to Bob, otherwise she sends
 classical information 1 to Bob. For example, if  Alice wants to send  Bob secret message 0100100  and
 Alice's measurement outcomes on the seven EPR pairs  are the states $|0\rangle,
 |1\rangle,|1\rangle,|0\rangle,|0\rangle,|0\rangle,|1\rangle$,  then Alice transfers the
 classical information 0010101 via classical public channel to Bob. Bob applies local measurement on his qubits $B$
 of the EPR
 pairs in the basis $\{|0\rangle_B, |1\rangle_B\}$. Clearly, he must obtain the same results $|0\rangle,
 |1\rangle,|1\rangle,|0\rangle,|0\rangle,|0\rangle,|1\rangle$  as  Alice. According to his measurement
 results and the classical information he received, Bob can infer the secret message 0100100 that Alice wants
 to transmit to him.

 It is clear that in our scheme the classical information resource is saved on since one bit classical information
 is only wanted to transmit one bit secret message.

 Evidently,  the security of this protocol
  is only determined by the quantum channel. In fact  a perfect quantum channel can be achieved by
   using the schemes testing the security of
 EPR pairs.  Therefore this scheme for quantum communication using
 EPR pairs and local measurement is absolutely reliable, deterministic and secure.

As mentioned in Ref. \cite {s26},  Eve can  obtain information if
the quantum channel is not the perfect EPR pairs.

   For example if Eve couples EPR pair with her probe in preparing EPR pair, and makes the quantum channel in the
following entangled state
\begin{equation}
|\Psi\rangle_{ABE}=\frac {1}{\sqrt
2}(|000\rangle_{ABE}+|111\rangle_{ABE}),
\end{equation}
then she is in the same position with the legitimate party Bob.
But this case can be ruled out after  Alice and Bob check the EPR
pairs by means of their  local measurement in the basis
$\{|0\rangle, |1\rangle\}$ or basis $\{\frac {1}{\sqrt
2}(|0\rangle+|1\rangle), \frac {1}{\sqrt
2}(|0\rangle-|1\rangle)\}$ randomly and comparing their
measurement results as stated in Ref. \cite {s26}. If Eve uses so
called entanglement pair method  to obtain information, she will
also be found by Alice and Bob's test, which was shown in Ref.
\cite {s26}. So in any case, as long as an eavesdropper exists,
she will be found and we can realize secure quantum communication.

\section{Summary}

We  give a  scheme for quantum  communication. The communication
is based on EPR pairs functioning as quantum channel and local
measurement.  After insuring the safety of the quantum channel,
Alice and Bob apply local measurement on the EPR pairs and Alice
broadcasts encoding classical information publicly. According to
the broadcast classical information and his measurement outcomes
on the EPR pairs Bob can infer the secret messages directly. On
one hand, transmitting one bit secret message only needs
transmitting one bit classical information, on the other hand,
there is not a transmission of the qubit which carries the secret
message between Alice and Bob, neither the communication can be
interrupted by Eve, nor the secret information is leaked to Eve.
Therefore our new protocol has high capacity and defends signal
against interference.

 \acknowledgments This work was supported by Hebei Natural Science Foundation of China under Grant No:
A2004000141 and Key Natural Science Foundation of Hebei Normal University.

\end{document}